\documentclass[floatfix,prd,lengthcheck,showpacs,amssymb,amsmath,amsfonts,aps,altaffilletter,nofootinbib,nopreprintnumbers,showpacsm]{revtex4-1}

\usepackage[utf8]{inputenc}
\usepackage[dvipsnames]{xcolor}
\usepackage{graphicx}
\usepackage{subcaption}
\usepackage{amsmath}
\usepackage{acronym}
\usepackage{multirow}
\usepackage{tabu}
\usepackage{import}
\usepackage{hyperref}
\usepackage[normalem]{ulem}
\usepackage[thinc]{esdiff}
\usepackage{caption}
\usepackage{lineno}

\hypersetup{
    colorlinks=true,
    linkcolor=blue,
    filecolor=magenta,      
    urlcolor=cyan,
}

\begin{document}

\title[]{Extracting SASI signatures from Gravitational Waves of Core-Collapse Supernovae using the Hilbert-Huang Transform}

\author{A. Veutro$^{1a,b}$, I. Di Palma$^{1a,b}$, A. Zegarelli$^{2}$}

\affiliation{$^{1a}$ Universit\`a di Roma  {\it{La Sapienza}}, I-00185 Roma, Italy}
\affiliation{$^{1b}$ INFN, Sezione di Roma, I-00185 Roma, Italy}
\affiliation{$^{2}$ Ruhr University Bochum, Faculty of Physics and Astronomy, Astronomical Institute (AIRUB), Universit\"{a}tsstra\ss e 150, 44801, Bochum, Germany}

% Please add ORCIDs here in case we submit to a journal which accepts them
% Gareth: 0000-0002-0355-5998
% Alex: 0000-0002-1850-4587
% Marton: 0000-0002-5354-5683

%\date{\today}

\begin{abstract}
Core collapse supernovae are among the most energetic astrophysical events in the Universe. Despite huge efforts on understanding the main ingredients triggering such explosions, we still lack of compelling evidences for the precise mechanism driving those phenomena. They are expected to produce gravitational waves due to asymmetric mass motions in the collapsing core, and emit in the meanwhile neutrinos as a result of the interactions in their high-density environment. The combination of these two cosmic messengers can provide a unique probe to study the inner engine of these processes and unveil the explosion mechanism. Among the possible detectable signature, standing accretion shock instabilities (SASI) are particularly relevant in this context as they establish a direct connection between gravitational wave emission and the outcoming neutrino flux. In this work, Hilbert-Huang transform is applied to a selected sample of 3D numerical simulations, with the aim of identifying SASI contribution and extract its instantaneous frequency. The performance of the method is evaluated in the context of Einstein Telescope.
\end{abstract}

\maketitle

\section{Introduction}

Core collapse supernovae (CCSNe) have been attracting the attention of human beings for centuries and despite the knowledge achieved in the last decades, the precise mechanism behind the engine of those explosion is still under debate. CCSNe are one of the most violent phenomena in the Universe and represent the cataclysmic deaths of massive stars, occurring when stars with a mass roughly above $8-10 M_\odot$ exhausts its nuclear fuel, leaving a central iron core. With fusion no longer able to counteract gravity, the core of the star collapses inwards within milliseconds, forming a proto-neutron star (PNS) and driving an outward shock wave, which travels through the in-falling material until it stalls at about 100 km, after having lost its energy via photo-dissociation of iron nuclei and neutrino cooling. According to the neutrino-driven mechanism \cite{Bethe:1990mw}, the stalled shock is revitalized through neutrinos streaming away from the PNS. If the energy deposited is not enough, the shock wave is not revived and the in-falling material continues to accrete through the shock wave, eventually leading to black hole formation.
Thanks to their nature, CCSNe are precious targets for multi-messenger astronomy, because, apart from their multi-wavelength electromagnetic emission, they release most of the PNS gravitational binding energy ($\sim 10^{53}$ erg) in the form of neutrinos ($\sim 99\%$) as it has been confirmed by the detection of MeV neutrinos from SN1987A \cite{Li:2023ulf}. Moreover, due to violent mass motions involved, aspherical asymmetries raise up and a strong energy release in form of Gravitational Wave (GW) is expected. 
GWs and neutrinos represent key tools to study this kind of phenomena, because contrary to the electromagnetic counterpart, which snapshots images of optically thin regions far away from the central engine, neutrinos and GWs can travel freely inside a dense matter environment and are expected to provide direct probes of the inner-workings of CCSNe. \newline
Despite ongoing computational challenges, multi-dimensional simulations of CCSNe are steadily converging and now offer credible predictions of expected GW signals. Across different research groups, simulations reveal consistent features in the time-frequency domain. One that is particularly interesting in this context is related to the standing accretion shock instabilities (SASI) \cite{Foglizzo_2006, Blondin_2003, 10.1093/mnras/stv1463}. SASI arises from a feedback mechanism between the shock and the PNS surface: acoustic waves perturb the spherical shock and the perturbed shock excites vorticity perturbations that advect downstream towards the newly formed PNS. Apart from playing a key role in facilitating the neutrino mechanism of CCSNe, SASI induces clearly modulations in the neutrino signals and produces a characteristic low frequency signature (around 100 Hz) in the GW emission that persist when SASI dominates over neutrino-driven convection \cite{PhysRevLett.111.121104}. The correlation between these two signatures has been investigated \cite{Kuroda_2017}, thus highlighting the importance of a simultaneous neutrino and GW detection of next CCSNe. Future SASI detections will indeed be a breakthrough in astrophysics, that will allow us to unveil the inner mechanisms triggering one of the most powerful phenomena in the Universe.\newline
%These findings highlight the crucial role that SASI will have in the future simultaneous detection of neutrinos and GWs from CCSNe.  [...] \newline
The current novel and exciting era of multi-messenger astronomy has established a wide and well organized network of telescopes that are able to communicate and collaborate each other. The next decade promises many upgrades on current facilities and the construction of new generation detectors that will allow us to observe the Universe with an unprecedented precision. In this context, supernovae serve as crucial laboratories to study the nuclear matter under extreme condition, and neutrinos and GWs represent a unique tool to unveil the precise mechanism of these phenomena and answer the puzzling question behind the nature of the processes that take part in reviving the shock and starting the explosion. Novel detection method and analysis tools are required, especially in the case of GW search of CCSNe which, being an unmodeled search, prevents us from using the matched filter technique \cite{Messick:2016aqy, hanna2020fast, Sachdev:2019vvd, usman2016pycbc, dal2014implementing, Allen:2005fk, nitz2017detecting} that is generally used to analyze GWs from compact binary coalescences (CBC). \newline 
In this context, the Hilbert-Huang transform (HHT) \cite{HUANG199659, Huang1998, Huang2003} is a very powerful tool which could be used to precisely describe the time-frequency properties of GW signal. It has been used for the first time in GW analyses in 2007 \cite{2007PhRvD..75f1101C} and, starting from there, its effectiveness has been demonstrated in studies involving various sources, from CBC \cite{2016PhRvD..93l3010K} to supernovae \cite{Hu_2022}. \newline

Following what was done in \cite{PhysRevD.104.084063}, here we apply the HHT to analyze GW data from CCSNe aiming to extract SASI-induced component. We validate the method on a set of CCSN multi-dimensional simulations, proving its robustness, and we test it in the context of Einstein Telescope (ET) \cite{ET-0007C-20}, a third generation GW detector. \newline
The paper structure is organized as it follows. In section \ref{hht}, the theory behind HHT is summarized; in section \ref{methodology}, the methodology applied in this work is described together with the dataset of numerical simulations employed; in section \ref{results}, we present the results obtained by applying this method to ET; in section \ref{conclusion}, we briefly discuss about possible future improvements.

\begin{table*}[t]
    \centering
    \renewcommand{\arraystretch}{1.7}
    \begin{tabular}{ |w{c}{4cm}|w{c}{2cm}|w{c}{3cm}|w{c}{2cm}|  }
    \hline
    Authors & Year & Model progenitor & ZAMS \\
    \hline
    Andresen et al. \cite{Andresen:2017} & 2016 & s27 & 27 M$_\odot$ \\
    Kuroda et al. \cite{Kuroda_2016} & 2016 & s15 SFHx & 15 M$_\odot$ \\
    Kuroda et al. \cite{Kuroda_2017} & 2017 & s15 SFHx & 15 M$_\odot$ \\
    O'Connor \& Couch \cite{O’Connor_2018} & 2018 & mesa20\_v\_LR & 20 M$_\odot$ \\
    Pan et al. \cite{Pan_2021} & 2020 &  & 40 M$_\odot$ \\
    Powell et al. \cite{10.1093/mnras/stab614} & 2021 & z100 SFHx & 100 M$_\odot$ \\
    \hline
    \end{tabular}
    \caption{Summary of multi-dimensional simulations of non-rotating progenitor models analyzed in this study. \textit{Year} is the year of publication; \textit{Name} is the simulation name containing the major info like equation of state ($SFHx$), metallicity ($s$ for solar metallicity and $z$ for zero metallicity) and other specific parameters (software employed $mesa$; velocity dependence in neutrino transport $v$; low resolution $LR$); \textit{ZAMS} is the Zero-Age Main-Sequence that refers to the mass of the progenitor.
    %The final column (‘SASI mode’) indicates which intrinsic mode function (IMF) contains the SASI signature. 
    For Pan et al. 2020 \cite{Pan_2021}, no specific \textit{Model progenitor} is provided, as the study reports only a single simulation.}
    \label{tab:1}
\end{table*}

\section{Hilbert-Huang Transform}
\label{hht}
\begin{figure}
    \centering
    \includegraphics[width=1\linewidth]{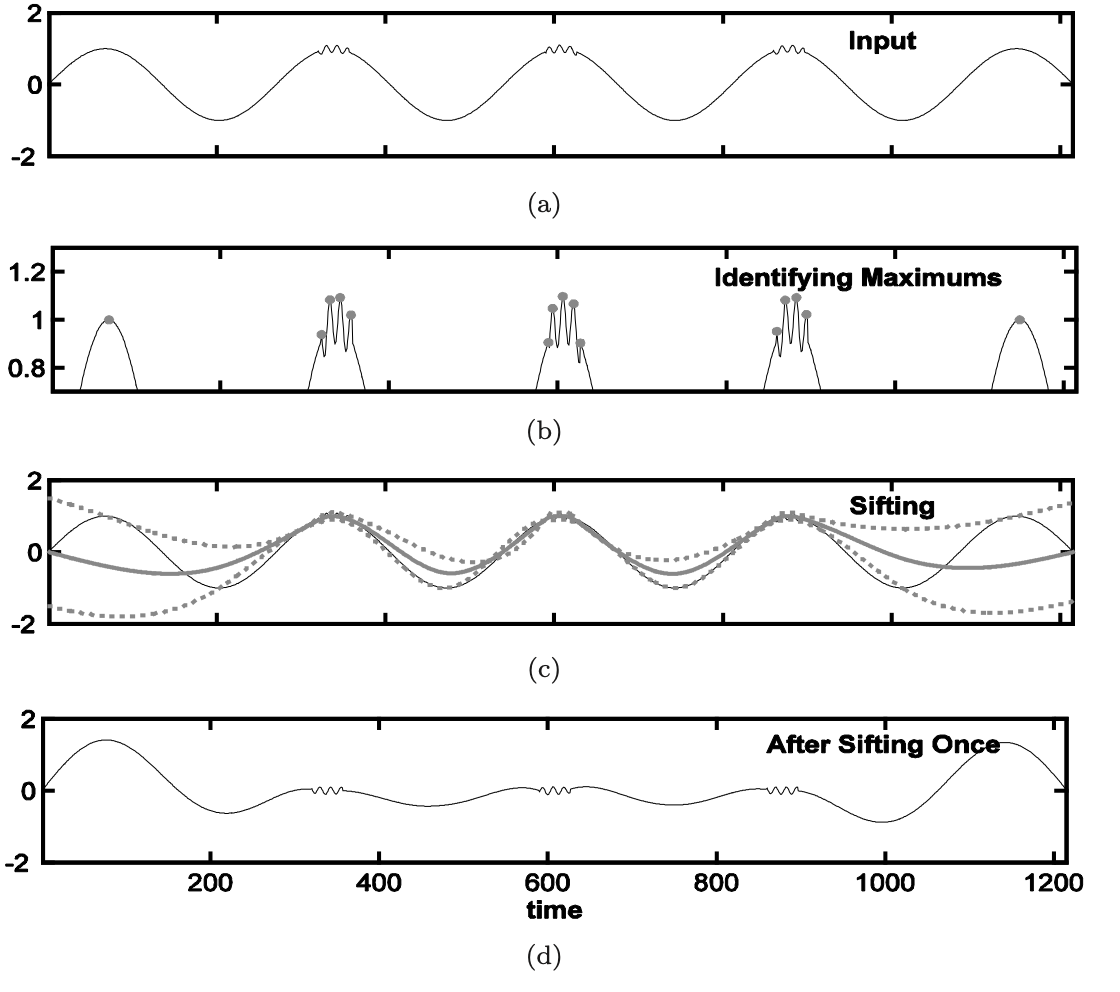}
    \caption{Taken from \cite{doi:10.1142/S1793536909000047}. IMF computation via EMD process. Panel (a) is the input; panel (b) identifies local maxima (gray dots); panel (c) plots the upper envelope (upper gray dashed line) and low envelope (lower gray dashed line) and their mean (bold gray line); and panel (d) is the difference between the input and the mean of the envelopes. The input data, which is composed of high-frequency intermittent oscillations riding on the fundamental low-frequency part, results in a first IMF that is a mixture of both components (mode mixing).}
    \label{mode-mixing}
\end{figure}

The HHT is an alternative time-frequency analysis algorithm that has been developed by Norden E. Huang et al. in 1998 \cite{Huang1998}. Compared to classical time-frequency analyses like spectrogram and wavelet analysis, the HHT does not rely on predefined basis function (adaptive approach) and its frequency resolution is not affected by the uncertainty principle, thus allowing the computed frequency to have the same time resolution as the original time series signal. \newline
HHT consists of two main components: empirical mode decomposition (EMD) and the Hilbert spectral analysis (HSA). EMD adaptively decomposes a signal $h(t)$ into a finite set of intrinsic mode functions (IMFs) $c_j(t)$, which represent simple oscillatory modes embedded within the original signal:
\begin{equation}
    h(t) = \sum\limits_{j=1}^N c_j(t) + r(t)
\end{equation}
Here, $r(t)$ is the residue left after extracting the IMFs. The resulting IMFs are simple oscillatory modes with time-varying amplitude and frequency, built to satisfy two key criteria: (1) the number of zero-crossings and extrema must either be equal or differ by at most one, and (2) at every point, the mean of the envelopes defined by the local maxima and minima must be zero.  These conditions ensure that IMFs represent well-behaved oscillatory modes with time-varying amplitude and frequency. \newline
Unlike predefined basis functions used in traditional methods, IMFs are data-driven and derived directly from the signal itself. Nevertheless, a fundamental limitation of standard EMD lies in its vulnerability to mode mixing \cite{2004RSPSA.460.1597W, doi:10.1142/S1793536909000047}, a phenomenon wherein a single IMF contains signals of widely disparate scales, or similar-scale components are distributed across multiple IMFs, leading to unclear physical meaning of individual IMF. Figure \ref{mode-mixing} give a brief explanation of this phenomena. To address this limitation, ensemble empirical mode decomposition (EEMD) was proposed \cite{doi:10.1142/S1793536909000047}. In EEMD, the original signal $h(t)$ is combined with white Gaussian noise realizations $N_i(t)$ to produce perturbed versions $s_i(t)$
\begin{equation}
    s_i(t) = h(t) + N_i(t) = \sum\limits_{j=1}^N c_{i,j}(t) + r_i(t)
\end{equation}
The final IMFs are obtained by averaging the decompositions over $I$ independent realizations:
\begin{equation}
    c_j(t) = \frac{\sum\limits_{i=1}^I c_{i,j}(t)}{I}
\end{equation}
Although adding noise may result in smaller signal-to-noise ratio, the added white noise will provide a relatively uniform reference scale distribution to facilitate EMD; therefore, the low signal–noise ratio does not affect the decomposition method but actually enhances it to avoid the mode mixing. \newline
Two key parameters characterize EEMD: the Gaussian noise amplitude $\sigma_{eemd}$ and the number of ensemble trials $N_{eemd}$. The value of $\sigma_{eemd}$ depends on the characteristics of the original signal: the larger the value, the more effective is the elimination of mode mixing \cite{Yuta:2011wht}. $N_{eemd}$ is primarily influenced by the value of $\sigma_{eemd}$. While an infinite number of trials would ideally eliminate the effect of added noise, using too many significantly increases computational cost. A detailed discussion on the final choice of $\sigma_{eemd}$ and $N_{eemd}$ used in this work is addressed in the next section. \newline
Once the signal is decomposed into IMFs, the Hilbert Transform is applied to each IMF to extract their instantaneous amplitude and frequency. This process enables the construction of the Hilbert Spectrum, a time-frequency-energy representation that reveals the evolution of oscillatory components over time. Given a real-valued function $f(t)$, the corresponding analytic signal $F(t)$ is defined as:
\begin{equation}
\label{hilbert-transform}
    v(t) = \mathcal{H}[f](t) = P \int _{-\infty} ^{+\infty} \frac{f(t')}{t-t'} dt' = f(t) * \left( \frac{1}{\pi t} \right)
\end{equation}
\begin{equation}
    F(t) = f(t) + i v(t) = a(t)e^{i \phi(t)}
\end{equation}
Here, $\mathcal{H}[f](t)$ denotes the Hilbert Transform of $h(t)$, $P$ represents the Cauchy principal value, and $*$ denotes convolution. From the analytic signal $F(t)$, we can define the following quantities: 
\begin{itemize}
    \item Instantaneous amplitude:
\begin{equation}
    IA(t) = | F(t) | = \sqrt{f(t)^2 + v(t)^2}
\end{equation}
    \item Instantaneous phase:
\begin{equation}
    \phi(t) = Arg[F(t)] = \tan^{-1} \left[ \frac{v(t)}{f(t)} \right]
\end{equation}
    \item Instantaneous frequency:
\begin{equation}
    IF(t) = \frac{1}{2 \pi} \diff{\phi(t)}{t}
    \label{if}
\end{equation}
\end{itemize}
Apart from providing a clever representation of the signal in the time-frequency plane, the main advantage of using HSA relies on the fact that the resulting $IF(t)$ and $IA(t)$ keep the same time resolution as the one of the input signal. \newline
In addition to the Hilbert Transform, other techniques such as generalized zero-crossing and direct quadrature methods have also been proposed for estimating instantaneous frequency \cite{doi:10.1142/S1793536909000096}. However, it is crucial to note that instantaneous frequency is well-defined only for signals that satisfy the IMF conditions—specifically, the signal must be symmetric about zero and have the number of extrema and zero-crossings either equal or differing by at most one \cite{Huang1998}.

\section{Methodology}
\label{methodology}

In this section, the methodology applied in this this work is presented: from the description of the dataset, which is built of three-dimensional simulation of non-rotating progenitors with strong SASI activity, to the pipeline used to apply HHT on our GW data to identify and extract SASI-induced feature.

\subsection{Dataset}

The waveforms selected for this study have been extracted from a set of three-dimensional simulations of non-rotating progenitors, which are expected to account for the vast majority of CCSN events - only about 1\% of such events exhibit signatures of rapid rotation, such as those associated with hypernovae or long gamma-ray bursts \cite{1992ApJ...392L...9D, 1976Ap&SS..41..287B, Akiyama_2003, nature1992}.
From this broader set, we selected only those simulations that display clear evidence of SASI activity in the GW emission. The corresponding waveforms are listed in Table \ref{tab:1}. \newline
The selected progenitors span a range of Zero-Age Main-Sequence (ZAMS) masses from 15 to 100~$M_\odot$. All the selected simulations assume solar metallicity, with the exception of Powell et al. 2021 \cite{10.1093/mnras/stab614}, which uses a zero-metallicity environment. To enhance the development of strong SASI activity, a soft equation of state (EoS) has been adopted in Kuroda et al. 2016 \cite{Kuroda_2016}, Kuroda et al. 2017 \cite{Kuroda_2017} and Powell et al. 2021 \cite{10.1093/mnras/stab614}: SFHx which has been developed by Steiner, Hempel and Fischer in 2013 \cite{2013ApJ...774...17S}. In fact, depending on the EoS stiffness, the bounce shock can be formed at larger radius. Consequently, the prompt shock stalls at smaller radius in the soft EoS model SFHx, thus setting a favorable condition for the SASI development due to the shorter advective–acoustic cycle. On the other hand, O\'Connor and Couch 2018 \cite{O’Connor_2018} perform a set of simulations employing a state-of-the-art $20 M_\odot$ progenitor generated using Modules
for Experiments in Stellar Astrophysics ($MESA$) software \cite{Paxton_2011}, and the SFHo equation of state \cite{2013ApJ...774...17S}. They perform eight 3D simulations varying the resolution and the velocity dependence in neutrino transport. Among them, we select $mesa20\_v\_LR$ which is a low-resolution ($LR$) simulation that incorporates neutrino transport with full velocity dependence ($v$). It is worth noting that in the simulation from Pan et al. 2020 \cite{Pan_2021}, the physical origin of the observed low-frequency GW component at around 200 Hz is not explicitly identified. While it may be attributed to higher-order or SASI mode, the authors do not provide a definitive interpretation. Nonetheless, we include this waveform in our dataset to demonstrate the capability of the HHT disentangle distinct physical modes from complex GW signals.

\begin{figure}
    \centering
    \includegraphics[width=1\linewidth]{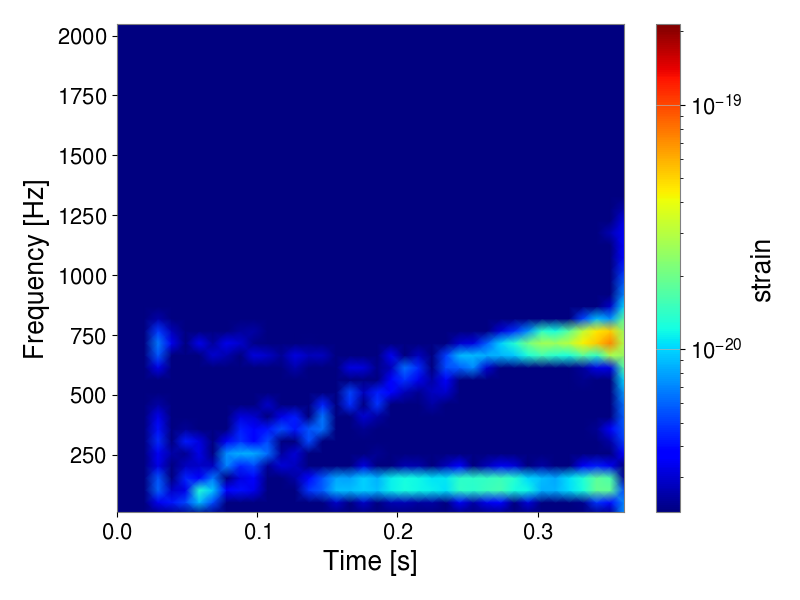}
    \caption{Kuroda et al. 2016 \cite{Kuroda_2016} plus polarization in time-frequency plane computed via short-time Fourier transform.}
    \label{kuroda2016}
\end{figure}

\begin{figure}
    \centering
    \includegraphics[width=0.9\linewidth]{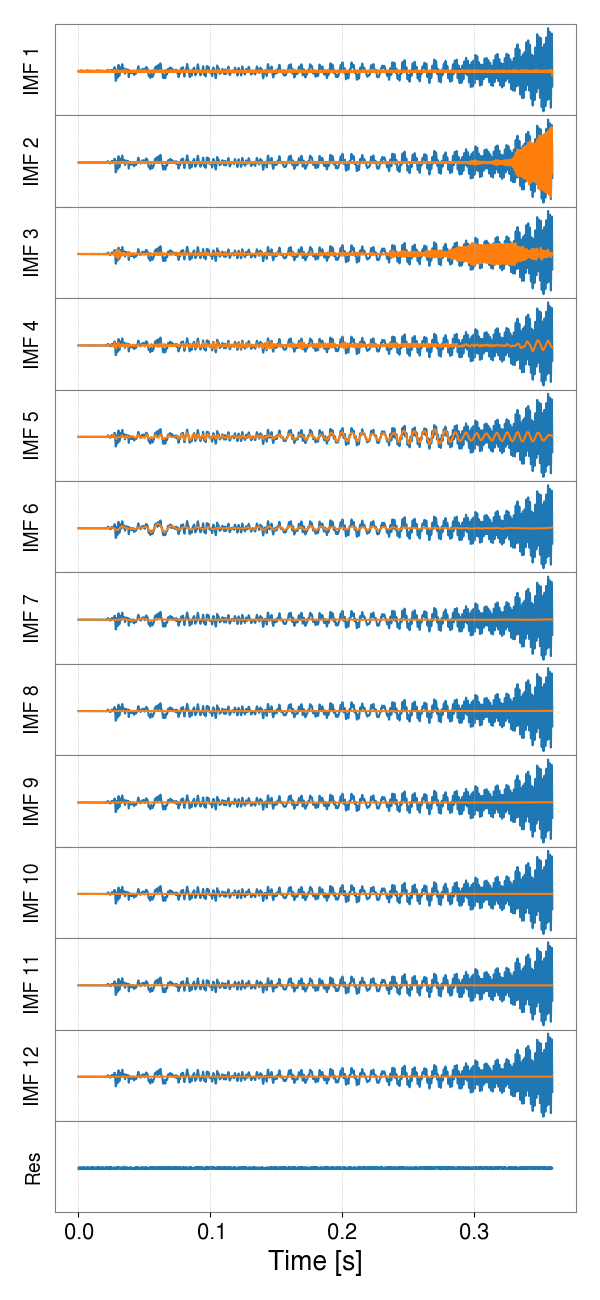}
    \caption{List of IMFs obtained by applying EEMD to the Kuroda et al. 2016 \cite{Kuroda_2016} signal. The blue curve shows the original input signal, while the orange curve represents the corresponding mode strain. The final panel displays the residual component from the decomposition.}
    \label{kuroda2016_imfs}
\end{figure}

\subsection{SASI extraction}

\begin{figure*}[htbp]
  \centering
  \includegraphics[width=1.2\linewidth]{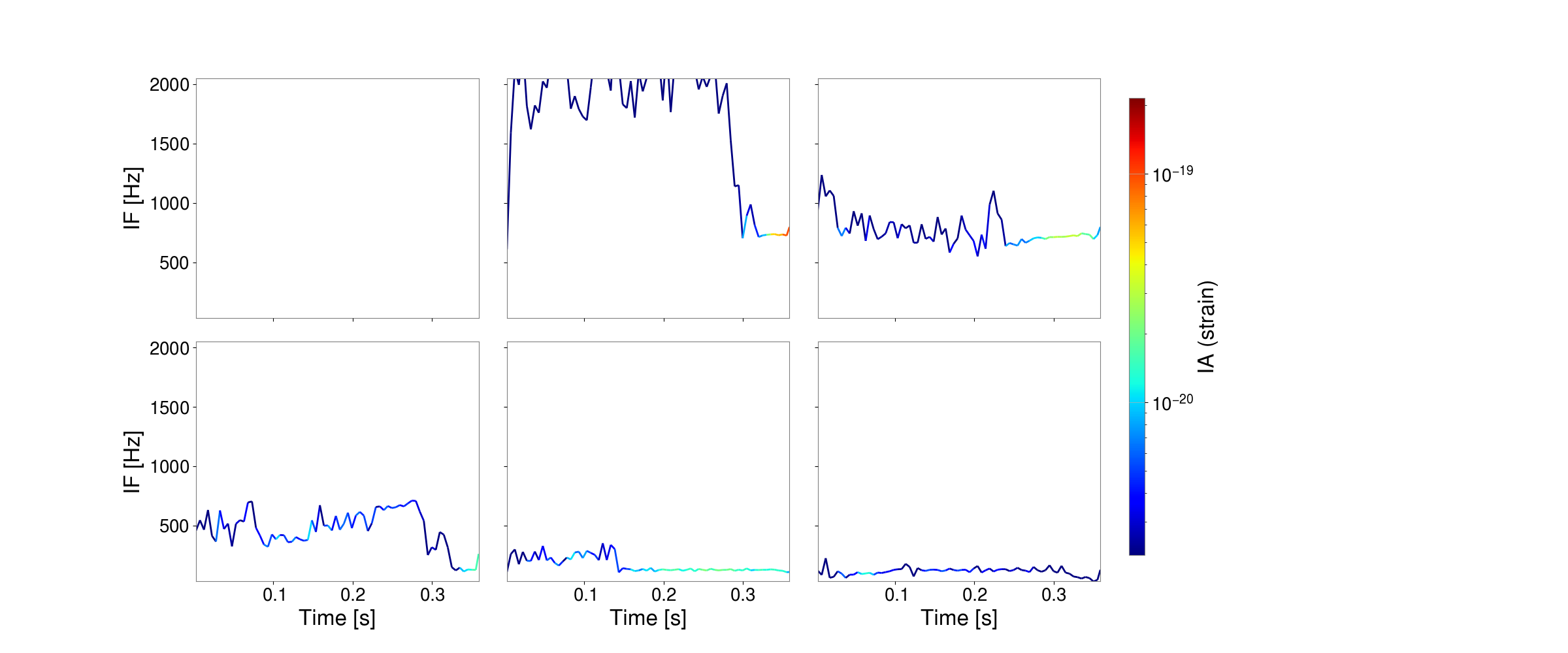}
    \caption{HHT-based time-frequency representations of the first six IMFs extracted from the Kuroda et al. 2016 simulation \cite{Kuroda_2016}. IF and IA are calculated using HSA, as described in Section \ref{hht}. To improve plot readability, the data have been downsampled to 200 Hz, following the approach in \cite{PhysRevD.104.084063}.}
    \label{kuroda2016_spec}
\end{figure*}

\begin{figure}
    \centering
    \includegraphics[width=1\linewidth]{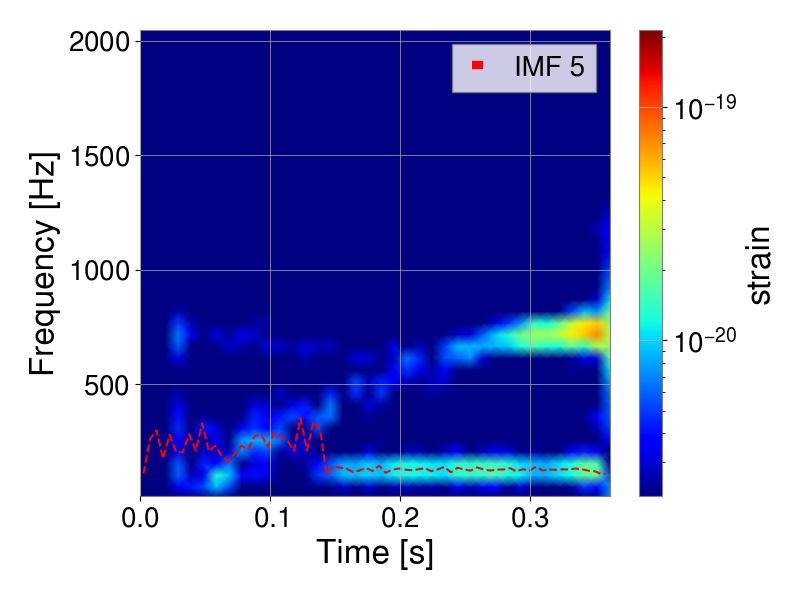}
    \caption{Superposition of Kuroda et al. 2016 \cite{Kuroda_2016} plus polarization in time-frequency plane and the 5th IMF (dotted red line), $i.e.$ the one containing SASI-induced feature.}
    \label{kuroda2016_sasi}
\end{figure}

\begin{figure*}[htbp]
    \centering
    \includegraphics[width=1.1\linewidth]{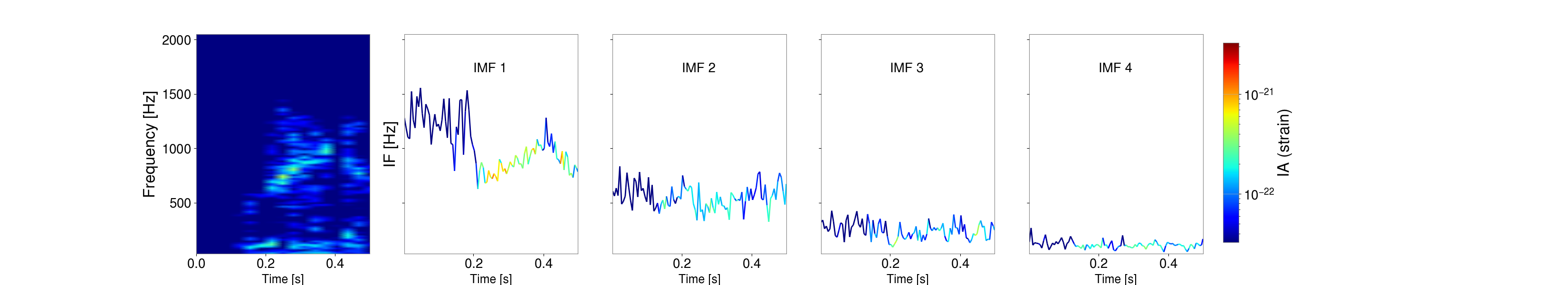}
    \includegraphics[width=1.1\linewidth]{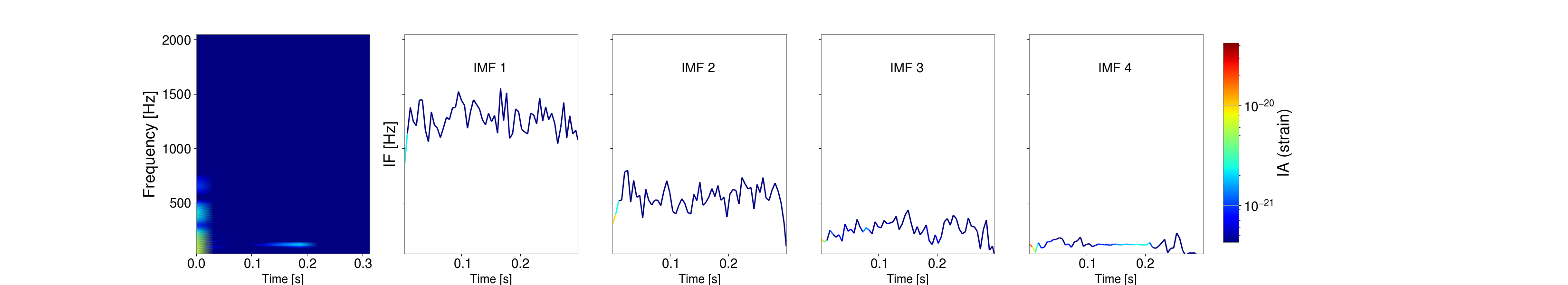}
    \includegraphics[width=1.1\linewidth]{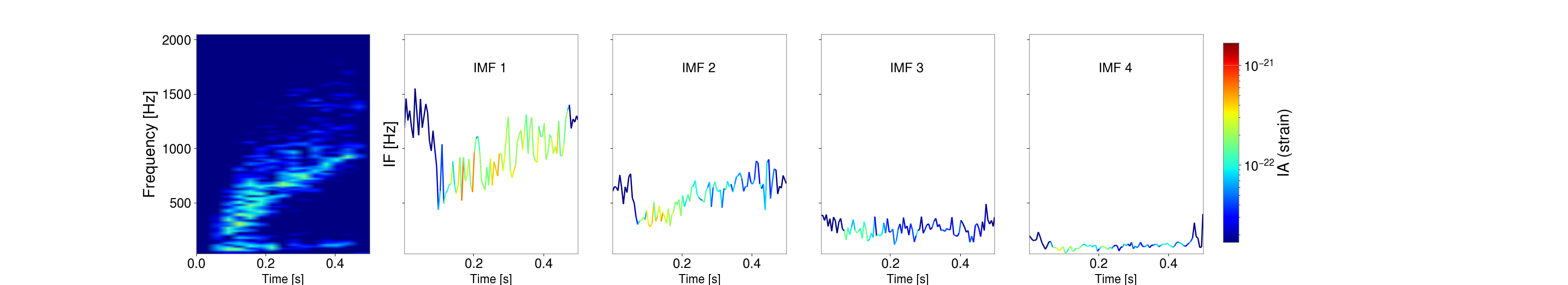}
    \includegraphics[width=1.1\linewidth]{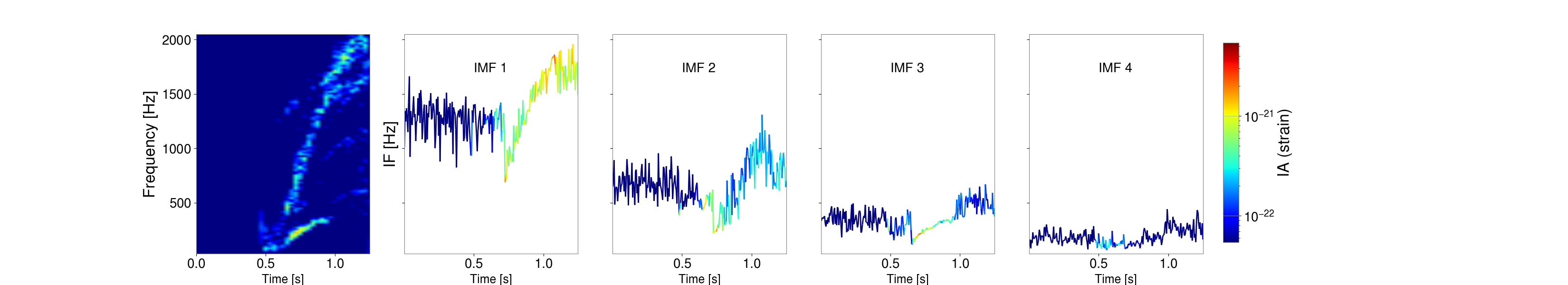}
    \includegraphics[width=1.1\linewidth]{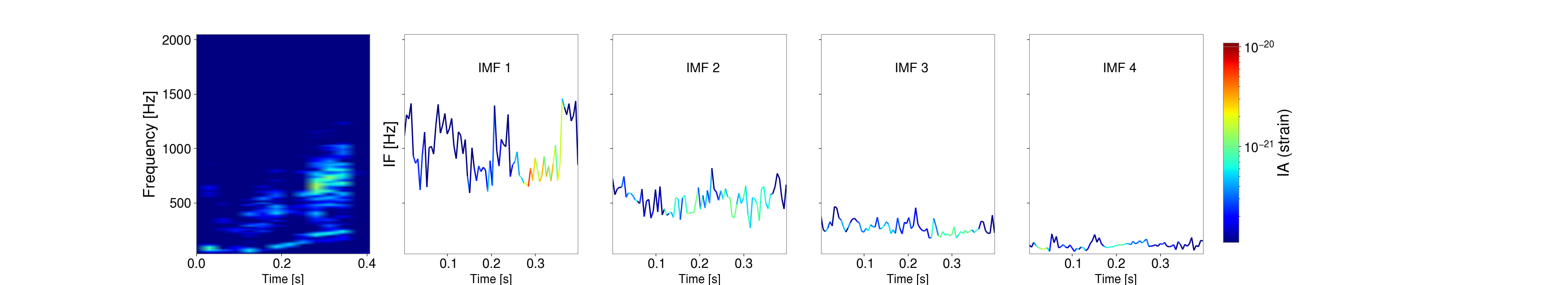}
    \caption{HHT-based time-frequency representations of the first four IMFs extracted from $h_+$ of the waveforms contained in our dataset (Table \ref{tab:1}). From top to bottom: (1) Andresen et al. 2016; (2) Kuroda et al. 2017; (3) O'Connor and Couch 2018; (4) Pan et al. 2020; (5) Powell et al. 2021.}
    \label{dataset_spec}
\end{figure*}

This work adopts HHT to extract the low-frequency SASI feature from CCSN GW signal. To illustrate the method in a straightforward manner, we use Kuroda et al. 2016 simulation \cite{Kuroda_2016} as example because of its loud SASI activity. Figure \ref{kuroda2016} shows the spectrogram of the input signal. This has been obtained using short-time Fourier transform. \newline
The first step of HHT consists in decomposing the input time series using EEMD. Time-series decomposition via EEMD has been implemented via the PyEMD Python package\footnote{The documentation is accessible here \url{https://pyemd.readthedocs.io/en/latest/intro.html}}. The decomposition has been conducted fixing $N_{eemd} = 100$, in order to keep a reasonable computational time (around 1 minute for a single decomposition), while $\sigma_{eemd}$ has been chosen as a trade-off between the fixed number of trials ($N_{eemd}$) and the quality of the reconstruction, which has been defined as the matched score $\eta$ between the original signal $h(t)$ and the reconstructed one $\tilde{h}(t)$, that is obtained by summing all the IMFs.
\begin{equation}
    \eta = \frac{(h|\tilde{h})}{\sqrt{(h|h)(\tilde{h}|\tilde{h})}}
    \label{match}
\end{equation}
Here, $(\cdot|\cdot)$ denoted the inner product without the noise weighting term. The value of $\sigma_{eemd} = 0.05 * |\max(h(t)) - \min(h(t))|$, ensures a reconstruction error below 0.3\%. \newline
Applying EEMD to the signal yields a set of IMFs, with the number of modes depending on the signal's complexity: more complex signals produce more IMFs. Figure \ref{kuroda2016_imfs} reports the 12 IMFs plus the residual obtained for our benchmark simulation. However, in this case, only the first six IMFs (excluding the first mode, which contains only high-frequency, low-amplitude noise) are found to carry significant physical information. \newline
After obtaining the IMFs, the HSA is applied to compute the instantaneous frequency of each mode (see Equation \ref{if}). Combining this with the instantaneous amplitude enables us to construct time-frequency representations for each IMF. Figure \ref{kuroda2016_spec} shows the spectrograms of the first six IMFs, down-sampled to 200 Hz for better visual clarity as it was done in \cite{PhysRevD.104.084063}. The SASI feature is clearly localized in the fifth IMF as it is shown in Fig. \ref{kuroda2016_sasi} which reports the superposition of input signal and the IF of IMF 5. Meanwhile, the most prominent structure in the original signal, a rising arch between 100 Hz and 750 Hz, is distributed across multiple IMFs (specifically, the second, third, and fourth modes). This "issue" has been previously discussed in \cite{doi:10.1142/S1793536909000047}, where it was noted that some physical features may require a combination of adjacent IMFs for accurate reconstruction, particularly when individual modes are not fully orthogonal. \newline
In Figure \ref{dataset_spec} we report the first four IMFs for all the waveforms used in this work. In Table \ref{tab:sasi} the results of SASI extraction is summarized.

\begin{table}[h]
\caption{Summary of SASI extraction results. \textit{SASI mode} column reports which IMF better fits the SASI-induced feature.}
\centering
\begin{tabular}{ll}
\hline
\textbf{Waveform} & \textbf{SASI mode} \\
\hline
Andresen et al. 2016     & 4th IMF \\
Kuroda et al. 2016       & 5th IMF \\
Kuroda et al. 2017       & 4th IMF \\
O\'Connor and Couch 2018 & 4th IMF \\
Pan et al. 2020          & 3rd IMF \\
Powell et al. 2021       & 4th IMF \\
\hline
\end{tabular}
\label{tab:sasi}
\end{table}

\section{Results}
\label{results}

\begin{figure*}[htbp]
  \centering
  \includegraphics[width=\linewidth]{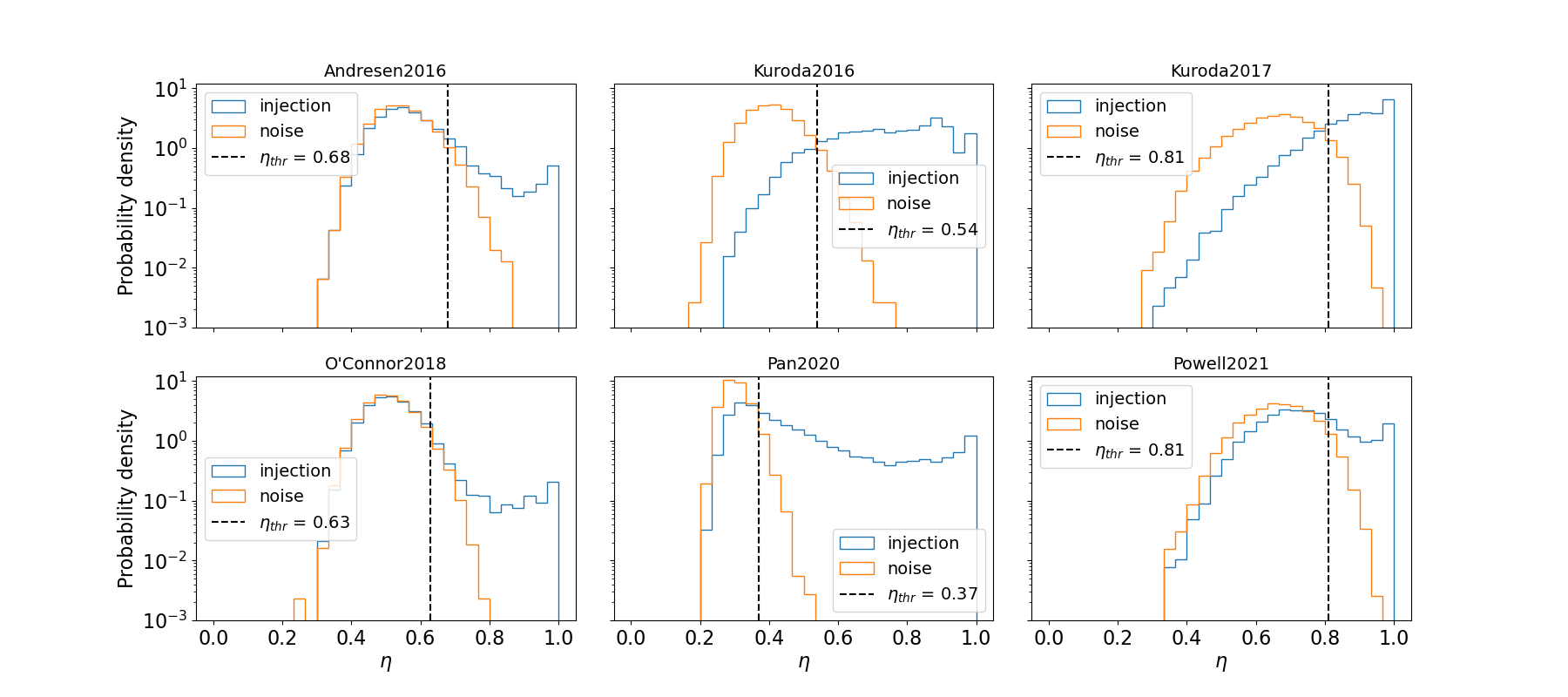}
    \caption{Match score $\eta$ distribution for the pure noise (orange) and injection (blue) samples for all the simulations considered in this work (see Table \ref{tab:1}). The injection has been performed with all the available waveforms and the source distance has been uniformly sampled up to 200 kpc. The black dotted line represents the threshold used to compute the detection efficiency. Its value is fixed by requiring the 95th percentile on the noise distribution.}
    \label{eff}
\end{figure*}

\begin{figure}
    \centering
    \includegraphics[width=\linewidth]{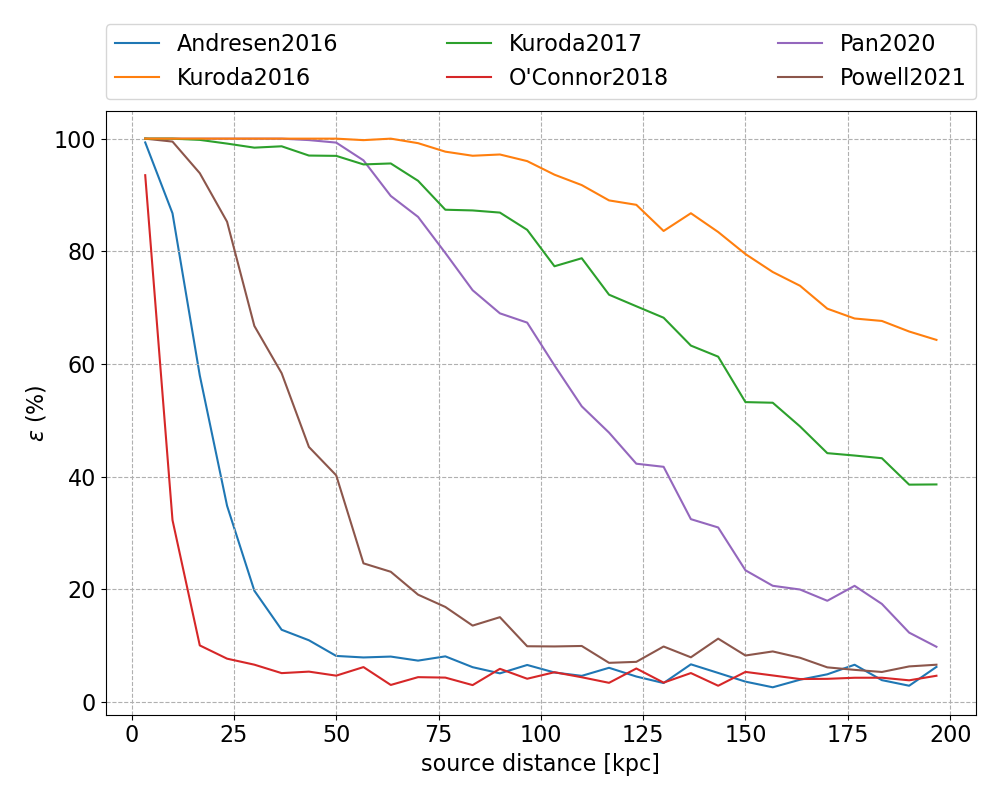}
    \caption{Detection efficiency $\epsilon$, computed with Equation \ref{epsilon}, in function of source distance. Different thresholds $\eta_{thr}$ have been applied to each of them depending on the $\eta_{bkg}$ distribution, $i.e.$ $\eta_{thr}$ is fixed by requiring the 95\% percentile on the noise distribution.}
    \label{eff_vs_dist}
\end{figure}

Inspired by what has been done in \cite{Yuan:2023umh}, we now present a method to compute the maximum source distance at which we can still reliably extract the SASI mode from data using the method described in the previous section. For this purpose, we focus on third-generation gravitational wave detectors, using the Einstein Telescope (ET) as our reference. Despite current uncertainties regarding the ET's exact geometry and location, we assume a configuration consisting of three interferometers with 10 km arms, arranged in a triangular layout with 60$^\circ$ opening angles \cite{Branchesi:2023mws}. A detailed analysis of how the results vary with different detector configurations is left for future work. \newline
We start by injecting the signal into simulated detector noise, Gaussian noise consistent with the sensitivity curve for 10 km interferometers reported in \cite{Branchesi:2023mws}. To this extent, we sample the source's sky location and distance, and we project the plus ($h_+$) and cross ($h_\times$) polarizations onto the detector frame using the antenna pattern functions $F_+$ and $F_\times$. The resulting signal strain $h(t)$ observed at detector is given by:
\begin{equation}
    h(t) = \frac{F_+(\theta,\phi)h_+ + F_\times(\theta,\phi)h_\times}{d}
\end{equation}
where $\theta$ and $\phi$ are the sky coordinates of the source, and $d$ is its distance. In this work, the signal injection assumes a uniform distribution of source distances up to 200 kpc and a uniform distribution over the sky. Due to the frequency-dependent nature of the detector’s power spectral density (PSD), it is essential to whiten the data to enhance signal visibility by suppressing frequency-dependent noise components. This whitening is performed using the \texttt{PyCBC} library \cite{alex_nitz_2024_10473621}, with a 50 second window around the injection used to reconstruct the PSD from data. \newline
%When including detector noise a pre-processing step is required. This involves estimating the source's sky location and distance to project the plus ($h_+$) and cross ($h_\times$) polarizations onto the detector frame using the antenna pattern functions $F_+$ and $F_\times$. The resulting detector strain h(t) is given by:
%\begin{equation}
 %   h(t) = \frac{F_+(\theta,\phi)h_+ + F_\times(\theta,\phi)h_\times}{d}
%\end{equation}
%where $\theta$ and $\phi$ are the sky coordinates of the source, and $d$ is its distance. Due to the frequency-dependent nature of the detector’s power spectral density (PSD), it is essential to whiten the data to enhance signal visibility by suppressing frequency-dependent noise components. This whitening is performed using the \texttt{PyCBC} library\footnote{Specifically, we use \texttt{pycbc.types.timeseries.TimeSeries.whiten} function.} \cite{alex_nitz_2024_10473621}, with a 50 second window around the injection used to estimate the PSD. \newline
%As described in the previous section, the first step involves injecting the signal into simulated detector noise. For clarity, we begin by presenting results based on the Kuroda et al. 2016 simulation, before extending the analysis to the entire dataset used in this study. 
%The detector noise is modeled as Gaussian noise consistent with the sensitivity curve for 10 km interferometers reported in \cite{Branchesi:2023mws}. The signal injection assumes a uniform distribution of source distances up to 200 kpc and a uniform distribution over the sky. 
We then apply the HHT to decompose both the injected signal and the noise-free source signal, the latter serving as our reference for assessing detection capability. Because the presence of noise increases the complexity of the resulting time series, IMFs may differ from those obtained in the noise-free case—additional modes may appear that reflect noise characteristics. To address this, we compute a match score $\eta_{inj}$ (as defined in Equation \ref{match}) between the SASI mode extracted from the signal at the source and each IMF from the signal injected in the noise. The IMF with the highest $\eta_{inj}$ is selected for further comparison. To minimize the influence of spurious components, the match score $\eta_{inj}$ is calculated only within the known time window of SASI activity. To assess detectability, we compare $\eta_{inj}$ against $\eta_{bkg}$, which is computed using the same procedure on noise-only data (no signal injected). A detection threshold $\eta_{thr}$ is then defined as the 95th percentile of the $\eta_{bkg}$ distribution—this represents the point below which the signal is no longer distinguishable from noise. Therefore, this naturally leads to the definition of a \textit{false alarm probability of reconstruction} of 5\%. The value of this threshold has been chosen accordingly to what was done in a similar analysis employing HHT for CCSN search \cite{Yuan:2023umh}.
Figure \ref{eff} shows the distributions of $\eta_{inj}$ and $\eta_{bkg}$ for a single ET interferometer. For the full ET network, the analysis accounts for each interferometer’s individual orientation and antenna patterns: after computing $\eta_{inj}$ for each, we retain the maximum score across the network.
Finally, Figure \ref{eff_vs_dist} presents the detection efficiency $\epsilon$ as a function of source distance, evaluated across the six numerical simulations used in this study (see Table \ref{tab:1}). The efficiency $\epsilon$ is computed as follows:
\begin{equation}
    \epsilon = \frac{N_{inj}(\eta_{inj}>\eta_{thr})}{N_{inj}}
    \label{epsilon}
\end{equation}
where $N_{inj}$ is the number of injections. \newline
Among the six simulations considered for this work, Kuroda et al. 2016 is the one achieving the best performance with and efficiency above 90\% at 100 kpc. On the other hand, O'Connor \& Couch has the lowest efficiency mainly due to the very low amplitude of the signal: almost 10 times weaker than the one of Kuroda 2016. 

\section{Conclusion}
\label{conclusion}

SASI play a key role in facilitating supernova explosion by enhancing the energy transfer to the medium. These instabilities leave a characteristic signature both in neutrino and GW emission. Specifically, in the latter it appears as a steady low frequency (around 100 Hz) mode. \newline
In this work we have explored a novel method developed to identify and extract GW signature induced by the SASI. The method is based on the use of HHT which adopts a mode decomposition algorithm, EEMD, and HSA to compute the instantaneous frequency of each individual mode without losing resolution with respect to the input signal. \newline
The method has been tested on a set of multi-dimensional simulations of non-rotating progenitors (about 99\% of CCSN events), in which SASI contribution appear to be strong enough. HHT allows a clever identification and extraction of the SASI mode for all of the signals considered here, despite their difference in frequency range and trend, $i.e.$ some of them manifest a monotonic increase of the SASI frequency mode. Moreover, for each of them, we computed the maximum distance at which the SASI mode can be extracted, taking into account the Einstein Telescope sensitivity with the triangle shape configuration, $i.e.$ three interferometers with a 60$^\circ$ opening angle and 10 km arms. Of course the maximum distance strongly depends on the simulation outcome that you are considering, thus spanning a wide variety of efficiency curve. Nevertheless, from our work, we can state that, in the best case scenario among the one considered here, $i.e.$ if we consider a CCSN similar to the one of Kuroda et al. 2016 (with SFHx equation of state), we should be able to identify SASI signature in GW data at a distance of 100 kpc with an efficiency above 90\%. This distance would allow us to efficiently detect eventual SASI feature from a CCSN exploding in the Large Magellanic Cloud at about 50 kpc from us. \newline
In future works, we aim to increase the numerical simulation dataset size and we plan to apply the method to real data from the Ligo-Virgo-Kagra collaboration. The feasibility of applying HHT also to neutrino light curve to extract SASI will be explored.

\section{Acknowledgements}
\label{s:acknow}

IDP acknowledges the support of the  Sapienza  School for Advanced Studies (SSAS) and the support of the Sapienza Grant No.RM124190B27960E5.

\bibliography{main}
\end{document}